\documentclass[conference]{IEEEtran}

\usepackage{cite}
\usepackage{amsmath,amssymb,amsfonts}
\usepackage{algorithmic}
\usepackage{graphicx}
\usepackage{textcomp}
\usepackage[table,xcdraw]{xcolor}
\usepackage{array}
\usepackage[colorlinks, citecolor=blue]{hyperref}
\usepackage{balance}
\usepackage[vlined,linesnumbered,ruled,boxed]{algorithm2e}
\usepackage{framed}
\usepackage{url, lscape, subfig, multirow, listings, verbatim, alltt, wrapfig, boxedminipage, bookmark}
\usepackage{epsfig}
\usepackage{xspace}

\newcommand{\comments}[1]{}

\begin{document}

\title{Exploring Challenges in Developing Edge-Cloud-Native Applications Across Multiple Business Domains}

\author{
\IEEEauthorblockN{Pawissanutt Lertpongrujikorn\IEEEauthorrefmark{1}, Hai Duc Nguyen\IEEEauthorrefmark{2}, Juahn Kwon\IEEEauthorrefmark{1}, Mohsen Amini Salehi\IEEEauthorrefmark{1}}
\IEEEauthorblockA{\IEEEauthorrefmark{1}High Performance Cloud Computing (HPCC) lab, University of North Texas, USA \\
Email: \{pawissanutt.lertpongrujikorn, juahn.kwon, mohsen.aminisalehi\}@unt.edu}
\IEEEauthorblockA{\IEEEauthorrefmark{2}Argonne National Laboratory and University of Chicago, USA \\
Email: hai.nguyen@anl.gov}
}

\maketitle

\begin{abstract}
As the convergence of cloud computing and advanced networking continues to reshape modern software development, edge-cloud-native paradigms have become essential for enabling scalable, resilient, and agile digital services that depend on high-performance, low-latency, and reliable communication. This study investigates the practical challenges of developing, deploying, and maintaining edge-cloud-native applications through in-depth interviews with professionals from diverse domains, including IT, finance, healthcare, education, and industry. Despite significant advancements in cloud technologies, practitioners—particularly those from non-technical backgrounds---continue to encounter substantial complexity stemming from fragmented toolchains, steep learning curves, and operational overhead of managing distributed networking and computing, ensuring consistent performance across hybrid environments, and navigating steep learning curves at the cloud-network boundary.  Across sectors, participants consistently prioritized productivity, Quality of Service, and usability over conventional concerns such as cost or migration. These findings highlight the need for operationally simplified, SLA-aware, and developer-friendly platforms that streamline the full application lifecycle. This study contributes a practice-informed perspective to support the alignment of edge-cloud-native systems with the realities and needs of modern enterprises, offering critical insights for the advancement of seamless cloud-network convergence.
\end{abstract}

\begin{IEEEkeywords}
edge-cloud-native, cloud computing, multi-domain, interviews, practical challenges
\end{IEEEkeywords}

\section{Introduction}
\label{sec:intro}

The convergence of cloud computing and high-speed networking has fundamentally reshaped modern business strategies by enabling unprecedented scalability, agility, and resilience \cite{10913359}. This evolution is pushing computation closer to the source of data, giving rise to distributed architectures like edge and fog computing. Consequently, there is a growing demand for models that support not just computation but also guaranteed low-latency responsiveness, localized decision-making, and flexible deployment across a hybrid cloud-to-edge continuum \cite{fatima2022production, modupe2024reviewing}. The underlying network fabric is no longer a simple transport layer but a critical, programmable component for service delivery. These capabilities are vital for several business domains—such as healthcare, IoT, finance, and robotics—where fast, adaptive, and network-aware software development and deployment are critical.

Despite significant advancements in cloud-native technologies, a persistent gap remains between the challenges acknowledged in academic research and those experienced by practitioners in real-world business domains \cite{ESASHI26}. Scholarly efforts often overlook critical constraints such as steep learning curves, fragmented toolchains, and limited technical capacity—especially prevalent among small enterprises and domain-specific organizations. These oversights are partly due to academic reliance on idealized assumptions, including underestimated cost implications, simplified system models, and minimal consideration of organizational limitations. This disconnect is further widened by the rapid evolution of cloud platforms and a lack of sustained collaboration between academic researchers and industry stakeholders. As a result, many proposed cloud-native frameworks, despite their conceptual promises, fall short when applied in operational environments.

To help bridge this research-practice divide, this paper presents preliminary empirical findings from interviews with domain-specific professionals directly engaged in the design, development, and adoption of cloud-native systems. These interviews aim to illuminate issues that are often underrepresented in academic discourse—specifically, the practical pain points experienced by practitioners, the expectations they hold for emerging technologies, and the business factors that shape decisions around cloud adoption and migration. By centering these real-world perspectives, this study seeks to surface latent challenges that hinder the practical uptake of academic innovations. By achieving this, we advocate for a more grounded and iterative research agenda—one that not only prioritizes technical novelty but also accounts for the operational, organizational, and economic realities of modern software development ecosystems.

This work is driven by the following hypothesis: 

\vspace{1mm}
\noindent\textbf{Hypothesis.} \textit{The current real-world practice of cloud-native application development, deployment, and maintenance is complicated, requiring significant time and specialized expertise, which increases overall costs.}

This hypothesis stems from a thorough examination of existing cloud-native systems and industry reports. Through our preliminary investigation, we identified a recurring pattern: domain-specific professionals often struggle with fragmented toolchains, complex deployment processes, and steep learning curves associated with distributed computing. 

To verify this hypothesis, we conducted interviews with 21 developers and industry professionals in multiple business domains, asking a set of questions that focused on the pain points they faced, the solutions or expectations for solutions, and the obstacles to implementing them.

The main contributions of this paper are as follows:

\begin{enumerate} 
    \item Identifying the real demands and challenges faced by industries in developing cloud-native applications through interviews (or surveys).
    \item Analyzing the interview results to extract key insights relevant to application development in practice.
    \item Proposing future research directions based on data-driven findings and practical needs.
\end{enumerate}

The rest of this paper is structured as follows. In Section~\ref{sec:motivation}, we present our motivation of this work. Section~\ref{sec:strategy} illustrates the strategies of interviews conducted. Each Section~\ref{sec:pain points},~\ref{sec:expectations}, and~\ref{sec:cost of migration} provides detailed data--pain points, expectations, and concerns in migration--driven from our interviews. In Section~\ref{sec:discussion}, we discuss our primary analyses. Section~\ref{sec:summary} summarizes this work with a brief suggestion for future works.

\section{Motivation}
\label{sec:motivation}

Cloud computing emerged in the early 2000s alongside advancements in virtualization and web hosting technologies \cite{aws_free_tier}. Virtualization tools like VMware \cite{vmware} enabled hardware abstraction, allowing multiple virtual machines (VMs) to run on a single physical server, thus greatly improving resource utilization. This abstraction laid a foundation for modern cloud platforms. By hiding infrastructure complexity, cloud promises to reduce infrastructure management burdens and empower organizations, especially those with limited technical capacity, to develop and scale digital services more easily. This vision led to widespread adoption of cloud-native technologies such as container orchestration systems \cite{k8s} and serverless computing \cite{aws_lambda, gcloud_func}.

Despite their potential, modern cloud platforms have evolved into a highly complex and fragmented landscape.
In fact, major commercial cloud providers (e.g., Amazon AWS \cite{aws_free_tier}, Google Cloud \cite{gcp}, and Microsoft Azure \cite{azure_enfunc}) have collectively introduced over \textbf{500} new services since \textbf{2006} \cite{borra2024comparison}. Many of these services demand deep technical expertise to configure, integrate, and operate effectively. Technologies like Kubernetes \cite{k8s}, Function-as-a-Service (FaaS) \cite{aws_lambda, gcloud_func}, CI/CD pipelines \cite{github_actions}, and microservices architectures compound steep learning curves and require developers to master a large and ever-growing set of tools and best practices. As a result, fundamental routine tasks--such as deploying an application or configuring its performance--now require a high level of expertise in cloud-native principles and platform-specific knowledge \cite{lertpongrujikorn2024streamlining}. 


This escalating technical and conceptual complexity directly contradicts the cloud's foundational promise of streamlined IT and universal accessibility. For startups, small enterprises, and domain-focused organizations without dedicated technical teams, adopting cloud-native solutions can be prohibitively complicated. Organizations in these categories often face challenges such as talent shortages, prolonged development cycles, debugging and integration overhead, and constrained budgets. Furthermore, the collaboration between domain experts and technical teams can be hindered by communication gaps and mismatched mental models. Translating domain-specific needs into robust cloud-native implementations is often non-trivial, further raising the barrier to entry.

To better understand the scope and nature of these challenges, we propose conducting structured interviews with professionals--including developers, solution architects, domain experts, and organizational decision-makers--involved in the design, development, deployment, and maintenance of cloud-native applications. These interviews aim to evaluate whether real-world cloud practices align with the hypothesis.

\section{Interview Methodology}
\label{sec:strategy}

To verify the hypothesis, we conducted \textbf{21} independent interviews with technology professionals from various domains. These interviews were designed to uncover practical challenges, assess tool usage, and understand the decision-making processes that influence the cost and complexity of cloud-native application development and deployment. Interview sessions were primarily held via Zoom between February and March 2025. Each session typically lasts 30 minutes, starting with a consistent set of core questions (see below) followed by an open-ended discussion to explore the unique experiences and expertise of each participant. To ensure transparency, consistency, and traceability, detailed records, including session recordings and interview notes, were maintained by two to three interviewers during each session.

\vspace{1mm}
\noindent\textbf{Domains}
To ensure broad representation, we selected participants from organizations in ten diverse sectors: agriculture, education, finance, healthcare, industrial, IoT, social network, hardware manufacturing, cybersecurity, and tech consultants. The sectors are chosen to reflect a wide range of system architecture, operational constraints, and domain-specific requirements. To simplify the analysis, we categorized organizations into two groups:
\begin{itemize}
    \item \textit{Tech} (38\% of interviews): organizations that provide digital solutions, such as cybersecurity services, technology consultancies, and hardware providers.
    \item \textit{Non-Tech} (62\%): other organizations whose core business is not technology-focused but still rely on software platforms and computation infrastructure (e.g., agriculture and healthcare).
\end{itemize}

\vspace{1mm}
\noindent\textbf{Participants and Their Roles:} 
We also diversify the set of participants across roles, including software developers, DevOps engineers, test engineers, startup entrepreneurs, and domain experts.
This mix allowed us to gather both hands-on technical perspectives and high-level strategic insights into how cloud-native technologies are adopted, maintained, and evolved within their organizations.

\vspace{1mm}
\noindent\textbf{Participant Demographics and Technology Landscape:} To provide a comprehensive view of our sample, Table~\ref{tab:demographics} summarizes key demographic characteristics and technology adoption patterns across the 21 participants.

\begin{table*}[ht]
\centering
\begin{tabular}{
|>{\centering\arraybackslash}p{5.5cm}
|>{\centering\arraybackslash}p{1.8cm}
|>{\centering\arraybackslash}p{2cm}
|}
\hline
\rowcolor[HTML]{C0C0C0}
\textbf{Characteristic} & \textbf{Count} & \textbf{Percentage} 
\\
\hline

\multicolumn{3}{|c|}{\textit{Organization Type}} 
\\
\hline
Tech Organizations & 8 & 38\% 
\\
Non-Tech Organizations & 13 & 62\% 
\\
\hline

\multicolumn{3}{|c|}{\textit{Business Domains}} 
\\
\hline
Education & 3 & 14\% 
\\
Finance & 3 & 14\% 
\\
Healthcare & 2 & 10\% 
\\
Industrial (Oil \& Gas, Manufacturing) & 3 & 14\% 
\\
IoT & 3 & 14\% 
\\
Agriculture & 2 & 10\% 
\\
Technology Consulting & 2 & 10\% 
\\
Cybersecurity & 1 & 5\% 
\\
Chips/Hardware & 2 & 10\% 
\\
Social Network & 1 & 5\% 
\\
\hline

\multicolumn{3}{|c|}{\textit{Infrastructure Types in Use (multiple possible)}} 
\\
\hline
Infrastructure-as-a-Service (IaaS) & 13 & 62\% 
\\
Platform-as-a-Service (PaaS) & 11 & 52\% 
\\
Function-as-a-Service (FaaS/Serverless) & 10 & 48\% 
\\
Edge Computing & 8 & 38\% 
\\
Hybrid Cloud (Edge + Cloud) & 6 & 29\% 
\\
Multi-Cloud & 4 & 19\% 
\\
Private Cloud & 3 & 14\% 
\\
AI-Specific Cloud & 2 & 10\% 
\\
\hline

\multicolumn{3}{|c|}{\textit{Organization Size}} 
\\
\hline
Startup/Small ($<$50 employees) & 4 & 19\% 
\\
Medium (50-500 employees) & 4 & 19\% 
\\
Large ($>$500 employees) & 7 & 33\% 
\\
Not Specified/Unknown & 6 & 29\% 
\\
\hline

\end{tabular}
\caption{Participant Demographics and Technology Adoption (N=21)}
\label{tab:demographics}
\end{table*}

The demographic distribution reveals several notable patterns. First, our sample achieved broad domain diversity, with representation across ten distinct business sectors. The 62\%/38\% split between non-tech and tech organizations aligns with our goal of capturing perspectives from domain-focused enterprises that may lack deep technical expertise alongside technology-oriented companies.

Regarding infrastructure adoption, traditional IaaS remains the most prevalent (62\%), reflecting the continued reliance on virtual machines and direct infrastructure control. However, managed services are gaining traction, with 52\% utilizing PaaS and 48\% adopting serverless/FaaS architectures. This suggests a gradual shift toward higher-level abstractions that reduce operational overhead.

Notably, 38\% of participants operate edge computing deployments, highlighting the growing importance of distributed architectures for latency-sensitive and network-constrained applications. Among these, 29\% employ hybrid edge-cloud configurations, indicating that pure edge or pure cloud solutions are often insufficient for real-world requirements. The presence of multi-cloud strategies (19\%) and private cloud deployments (14\%) further underscores the heterogeneity and complexity of modern cloud-native landscapes.

Organization size distribution shows that one-third of participants represent large enterprises ($>$500 employees), while nearly 40\% come from small to medium organizations. This mix provides insights into how resource constraints and organizational capacity influence cloud adoption strategies and pain points.

Having established the demographic composition of our sample, we now turn to the interview execution. The semi-structured interviews were conducted using a standardized protocol to ensure consistency while allowing for open-ended exploration of emerging topics.

\vspace{1mm}
\noindent\textbf{Core Interview Questions}
Each interview began with a consistent set of guiding questions to collect comparable data, while still allowing flexibility for domain-specific elaboration. Main areas of focus include:
\begin{itemize}
    \item \textit{Technology.} What technologies are currently in use, and what alternatives (if any) are under consideration and evaluation?
    \item \textit{Desired Features:}
    What capabilities would participants want from the underlying technologies to better support their needs?
    \item \textit{Pain Points:}
    What are the most noticeable technical or organizational challenges they face in development, deployment, and maintenance?
    \item \textit{Addressing Pain Points} Are the pain points solved? Would participants consider switching to a new platform as a part of the solution? What technical, cost-related, or organizational barriers hinder such transitions?
\end{itemize}

In the following sections, we present our findings based on participants' responses to these questions. However, before proceeding, it is essential to note that the organization of themes in our data presentation does not imply their relative importance across domains. For instance, if cost is explicitly discussed by certain organizations but not others, this does not imply that cost is irrelevant or overlooked—it may simply not have emerged as a dominant theme in those particular interviews. Furthermore, for various reasons (primarily personal or business-related), participants are not required to answer all questions. As a result, any unanswered responses are marked as unknown and excluded from the analysis.

\section{Pain Points of Current Practice}
\label{sec:pain points}

\begin{table*}[ht]
\vspace{-4mm}
\centering
\begin{tabular}{
|>{\centering\arraybackslash}p{2.2cm}
|>{\centering\arraybackslash}p{6cm}
|>{\centering\arraybackslash}p{2.3cm}
|>{\centering\arraybackslash}p{2.6cm}
|}
\hline
\rowcolor[HTML]{C0C0C0}
\textbf{Category} &
\textbf{Pain point} &
\textbf{Business domain} &
\textbf{Resolved} 
\\
\hline

Cost &
High cost for usage of cloud services and significant effort for optimizing resource utilization &
Non-Tech(40\%) Tech(60\%) Total(24\%) &
Partly(20\%)   Unknown(80\%)
\\
\hline

Operational maintainability &
Lack of dedicated technical personnel, delayed support of cloud service providers, lack of cloud support for legacy programs, and complex cloud-based business service configurations &
Non-Tech(67\%) Tech(33\%) Total(57\%) &
Resolved(33\%) Partly(33\%)  Unknown(33\%)
\\
\hline

Development complexity &
Lack of dedicated technical personnel, distributed system across Edge-Cloud, and hybrid data flows (on-prem to cloud) &
Non-Tech(80\%) Tech(20\%) Total(48\%) &
Resolved(10\%) Partly(30\%) Unresolved(20\%)  Unknown(40\%)
\\
\hline

Responsiveness &
Database bottleneck and lack of architectural support for data locality &
Non-Tech(56\%) Tech(44\%) Total(43\%) &
Resolved(33\%) Unknown(67\%)
\\
\hline

Reliability \& Availability &
 No SRE or full-time DevOps, unreliable network connection in edge systems, and system misconfigurations &
Non-Tech(57\%) Tech(43\%) Total(33\%) &
Resolved(0\%) Partly(57\%)  Unknown(43\%)
\\
\hline

Security \& Privacy &
Authentication on edge with unreliable network, handle sensitive data, and government regulations (e.g., PDPA)&
Non-Tech(75\%) Tech(25\%) Total(38\%) &
Resolved(37.5\%) Partly(25\%)  Unknown(37.5\%)
\\
\hline

\end{tabular}
\caption{Pain Points Across Infrastructure and Business Domains}
\label{tab:painpoints_all}
\end{table*}

During interviews, participants described a wide range of challenges encountered when adopting and scaling cloud-based solutions. These challenges span performance bottlenecks, DevOps limitations, infrastructure trade-offs (particularly in hybrid and edge-cloud settings). Based on their responses, we classified the reported issues into six distinct categories. Table~\ref{tab:painpoints_all} summarizes these pain point categories, including representative examples, the total and per-domain percentage of participants who explicitly described each issue, and whether the issue had been resolved at the time of the interview. Key findings and trends are highlighted below.

\vspace{1mm}
\noindent\textbf{Most pain points stem from technology-related categories}: \textit{Operational maintainability} emerged as the most frequently reported pain point, affecting 57\% of all participants. This issue was often linked to a lack of dedicated technical personnel, limited cloud provider support, and the complexity of configuring and managing cloud services. Thus, it is understandable that the second and third most challenging pain points belong to \textit{development complexity} (48\%) and \textit{responsiveness} (43\%), which also require strong technical expertise to address. On the other hand, \textit{cost} appears to be of less concern, which may seem counterintuitive. This can be attributed to the fact that some interviewees, particularly those in technical or operational roles, may not have emphasized this issue due to their limited involvement in budgeting or financial decision-making.

\vspace{1mm}
\noindent\textbf{Non-tech participants face significantly more challenges than tech participants.}. Across all domains, the percentage of non-tech participants experiencing pain points is mostly consistently higher than that of their tech counterparts. Noticeably, the gap is $4\times$ and $3\times$ in \textit{development complexity} and \textit{security and privacy}, respectively. This trend suggests that a lack of technical background can substantially hinder the effective adoption and use of cloud services.

\vspace{1mm}
\noindent\textbf{Most pain points remain unresolved} As shown in Table~\ref{tab:painpoints_all}, fewer than half of the reported pain points had been fully resolved at the time of the interview (excluding cases where responses were withheld due to business sensitivity). In particular, \textit{reliability and availability} concerns were mostly reported by organizations lacking dedicated DevOps or SRE teams. None of these organizations reported full resolution of the issue, though some had implemented partial mitigations.

\subsection{Illustrative Examples from Interviews}

To provide concrete context for the pain points identified in Table~\ref{tab:painpoints_all}, we present representative cases synthesized from our interview data. These vignettes illustrate how multiple challenges often intersect in real-world deployments.

\vspace{2mm}
\noindent\textbf{Case 1: Healthcare Startup---Operational and Reliability Challenges}

A small healthcare technology company (fewer than 50 employees) operates a cloud-based platform combining serverless and traditional infrastructure. The organization reported persistent challenges with notification delivery reliability across different mobile platforms, unplanned database maintenance windows causing service interruptions, and frequent network connectivity issues. Without dedicated DevOps personnel, incident response required several hours of manual log analysis to identify affected services. While the organization addressed some issues through database redundancy and automated monitoring, the fundamental resource constraint of lacking full-time operational staff remained unresolved.

This case exemplifies the intersection of \textit{reliability \& availability} (insufficient operational staffing), \textit{operational maintainability} (dependency on cloud provider support responsiveness), and \textit{responsiveness} (database availability gaps). These challenges are characteristic of resource-constrained small enterprises attempting to maintain production systems without dedicated site reliability engineering teams.

\vspace{2mm}
\noindent\textbf{Case 2: Industrial Sector---Edge-Cloud Hybrid Complexity}

An industrial automation company deploying hybrid edge-cloud architectures for remote equipment monitoring faced challenges specific to distributed environments with unreliable connectivity. The organization needed authentication mechanisms that function during extended network outages, immediate credential revocation capabilities despite connectivity limitations, and protection against physical device compromise in remote locations. Additionally, high-volume telemetry data generation (approximately 1 MB per second per device) created data synchronization and deduplication challenges when network partitions occurred.

This vignette highlights \textit{security \& privacy} concerns (offline authentication and physical security), \textit{reliability} issues (intermittent network connectivity), and \textit{development complexity} (conflict-free data synchronization). Standard cloud platforms lacked native support for authentication systems resilient to extended disconnection periods and sophisticated data synchronization mechanisms required for these use cases.

\vspace{2mm}
\noindent\textbf{Case 3: Education Technology---Serverless Scalability Gaps}

An education platform leveraging serverless computing for personalized student interactions encountered architectural limitations when function-level auto-scaling did not extend to the database tier. Despite compute resources scaling automatically based on demand, database throughput became a bottleneck during peak usage. The organization implemented custom synchronization techniques and serialization strategies to work around these limitations, significantly increasing development effort. The absence of in-memory processing support in the serverless environment further complicated state management requirements.

This case illustrates \textit{responsiveness} challenges stemming from the mismatch between compute-layer and data-layer scaling capabilities. Organizations adopting serverless architectures must often design custom solutions to bridge gaps where cloud platform abstractions fail to provide end-to-end scalability guarantees.

\vspace{2mm}
\noindent\textbf{Case 4: Financial Institution---Hybrid Data Residency Requirements}

A large financial services organization maintains strict data residency policies where all sensitive data remains on-premises, traveling to public cloud environments only in encrypted form for processing. While this hybrid architecture satisfies regulatory requirements, it introduces substantial \textit{development complexity} through the proliferation of configuration parameters across container orchestration platforms and serverless function environments. The organization's structure further complicated operations, as development teams assumed DevOps responsibilities without dedicated operational support staff.

Despite the technical sophistication of the institution, the hybrid architecture imposed operational overhead exceeding that of pure cloud deployments. This demonstrates how regulatory compliance requirements can mandate architectural patterns that increase rather than reduce system complexity.

\vspace{2mm}
\noindent\textbf{Case 5: Government Education---Policy and Legacy Constraints}

A government educational institution operating private cloud infrastructure faced barriers to public cloud adoption stemming from data privacy regulations requiring extensive certification processes. Beyond the high cost of compliance certification, organizational structure created additional challenges: infrastructure teams and application development teams operated independently with limited communication, reflecting legacy organizational boundaries. Furthermore, contractual procurement cycles constrained technology migration decisions, as platform changes could only occur upon hardware contract expiration, regardless of technical merit.

This example underscores \textit{cost} concerns (regulatory compliance overhead), \textit{organizational constraints} (team silos), and \textit{operational maintainability} challenges (infrastructure-application integration gaps). The rigid procurement cycles illustrate how non-technical factors can dominate platform adoption timelines in public sector contexts.

\vspace{2mm}
These cases demonstrate that pain points rarely occur in isolation. Organizations typically face overlapping challenges across multiple categories, with resource constraints, regulatory requirements, and architectural decisions compounding the complexity of cloud-native adoption.

\vspace{2mm}
\noindent
\colorbox{blue!10}{
\parbox{0.96\linewidth}{
\underline{\textbf{Takeaway (Pain Point)}:} \emph{
Most reported pain points remain unresolved, especially for non-tech organizations lacking technical expertise and dedicated support teams.
}}}

\section{Expectations to Cloud Provider/Platform}
\label{sec:expectations}

\begin{table*}[ht]
\vspace{-4mm}
\centering
\begin{tabular}{
|>{\centering\arraybackslash}p{2.4cm}
|>{\centering\arraybackslash}p{7cm}
|>{\centering\arraybackslash}p{1.4cm}
|>{\centering\arraybackslash}p{0.7cm}
|>{\centering\arraybackslash}p{0.8cm}
|}
\hline
\rowcolor[HTML]{C0C0C0}
\textbf{Category} &
\textbf{Expectation} &
\textbf{Non-tech} &
\textbf{Tech} &
\textbf{Total} 
\\
\hline

Availability \& Reliability &
High availability (minimum 99.9\%), quick support contact, and well-defined disaster recovery &
67\% & 33\% & 43\%
\\
\hline

Security \& Privacy &
Strong and resilient security system with data encryption &
67\% & 33\% & 29\%
\\
\hline

Service Quality Assurance &
High SLA offering, high-level and rich configurability for performance optimization, system scalability, and capability &
57\% & 43\% & 33\%
\\
\hline

Unified Maintainability &
Support Hybrid cloud, technical support for automation (DevOps), and log aggregation &
43\% & 57\% & 33\%
\\
\hline

Programmability &
Ease of development and user-friendly interface &
50\% & 50\% & 38\%
\\
\hline

\end{tabular}
\caption{Interviewee Expectations Mapped to Infrastructure and Business Domains}
\label{tab:expectations_all}
\end{table*}

Table~\ref{tab:expectations_all} summarizes the key expectations that participants believe cloud platforms or other underlying technologies must fulfill to address their current pain points and better support application development, deployment, and management in the future.
We categorize these expectations into five feature groups, each representing a capability that cloud platforms should provide (see the \textit{Expectation} column). For example, \textit{availability and reliability} present the ability of a platform to maintain smooth application operation in the face of failures or disasters (e.g., large-scale outages). For each category, we report the percentage of non-tech, tech, and total participants who expressed a need for that support. Key findings include.

\vspace{1mm}
\noindent\textbf{Tech participants prioritize quality-of-service supports.}
A majority of tech participants emphasized the importance of performance-related guarantees, including \textit{availability and reliability} (67\%), \textit{security and privacy} (67\%), and \textit{service quality assurance} (57\%). This reflects the fact that their solutions are typically digital services. These services highly rely on the performance of underlying technologies (e.g., up-time, throughput, and latency) to deliver high-quality services and maintain a positive user experience, which is critical for user retention and business success. 

\vspace{1mm}
\noindent\textbf{Non-tech participants prioritize simplification and usability.} In contrast, non-tech participants more often highlighted the need for simplifying platform complexities. As discussed in previous sections, a lack of technical expertise is a major barrier for these users. Consequently, capabilities that reduce operational complexity or support no-code/minimal-effort development, such as \textit{unified maintainability} (57\%) and \textit{programmability} (50\%), were most frequently mentioned in this group.

\vspace{1mm}
\noindent\textbf{A broad range of support is needed across domains.}
When aggregating responses across both tech and non-tech participants, we observe a relatively even distribution of expectations across all five categories. This indicates a strong demand for comprehensive and unified platform support. This suggests that future cloud solutions should not only deliver high performance but also ensure accessibility and ease of use to serve a broad range of applications and users.

\vspace{2mm}
\noindent
\colorbox{blue!10}{
\parbox{0.96\linewidth}{
\underline{\textbf{Takeaway (Expectation)}:} \emph{
Future cloud platforms must deliver quality-of-service and simplicity to effectively support both tech and non-tech users.
}}}

\section{Migration Readiness}
\label{sec:cost of migration}

\begin{table*}
\centering
\begin{tabular}{
|>{\centering\arraybackslash}p{2.1cm}
|>{\centering\arraybackslash}p{7cm}
|>{\centering\arraybackslash}p{1.4cm}
|>{\centering\arraybackslash}p{0.7cm}
|>{\centering\arraybackslash}p{0.8cm}
|}
\hline
\rowcolor[HTML]{C0C0C0}
\textbf{Category} &
\textbf{Migration concern} &
\textbf{Non-tech} &
\textbf{Tech} &
\textbf{Total} 
\\
\hline

Cost &
Difficulty in estimating migration costs and high upfront migration cost &
67\% & 33\% & 14\%
\\
\hline

Technical Difficulty &
Software dependency on the current cloud platform, limited technical proficiency in migration, and uninterrupted SLA compliance during migration &
75\% & 25\% & 38\%
\\
\hline

Organizational Constraints &
Constraints due to existing contractual obligations with the current platform, and data policy-related legal constraints & 
67\% & 33\% & 14\%
\\
\hline

Developer Efforts &
Intensive manual migration efforts, time-consuming migration process, and potential need for code rewriting &
60\% & 40\% & 24\%
\\
\hline

Internal Resistance &
Greater familiarity with the existing platform, and 
Employee feeling unsafe about adopting a new solution &
100\% & 0\% & 5\%
\\
\hline
\end{tabular}
\caption{Migration obstacles across business domains}
\label{tab:migration_concerns}
\end{table*}

We next investigate participants’ willingness to migrate their solutions to new underlying technologies or cloud platforms (e.g., shifting from container-based services~\cite{ums23,fastmig24} to Function-as-a-Service) to fully resolve their pain points or improve the quality of their solutions. The questions arise from the rapid evolution of cloud technologies and the lack of standard service abstraction. These factors make service migration a significant undertaking—one that can yield long-term improvements but may also introduce short-term disruptions or risks.
When asked about migration readiness, participants expressed a \textbf{mixed reaction}. While there is general openness to migration, real-world adoption is tempered by five critical obstacles as shown in Table~\ref{tab:migration_concerns}.

\vspace{1mm}
\noindent\textbf{Many are open to migration.}
Across all reported obstacles, only a minority of participants were against migration. The most frequently raised concern, \textit{technical difficulty}, was mentioned by 38\% of participants. All other obstacles were mentioned by less than 30\% of participants. Notably, \textit{internal resistance} (e.g., fear of disruption or layoffs) was rarely identified as a primary deterrent, cited by only 5\%. These findings reflect a general recognition that migration offers clear \textit{long-term} benefits, such as performance improvements, cost optimization, and access to mature services, that often outweigh the \textit{short-term} costs and effort.

\vspace{1mm}
\noindent\textbf{Non-tech participants are more hesitant to migrate.} For instance, cost-related concerns among non-tech participants were reported at double the rate of their tech counterparts. It is worse for \textit{internal resistance} where \textit{all} objections are exclusively from non-tech participants. This trend aligns with earlier findings: without a strong technical background, non-tech participants tend to be more risk-averse when facing potentially disruptive changes. In contrast, tech participants are generally more confident in evaluating trade-offs and are thus more willing to pursue migration when justified by long-term gains.

\vspace{2mm}
\noindent
\colorbox{blue!10}{
\parbox{0.96\linewidth}{
\underline{\textbf{Takeaway (Migration Readiness)}:} \emph{
Most participants are open to migration, but non-tech users remain cautious due to concerns about cost, complexity, and the fear of disruption.
}}}

\section{Discussion}
\label{sec:discussion}



\subsection{Comprehensive Insights}

We provided detailed interpretations and takeaways from the data tables—covering pain points, expectations, and migration concerns—in Sections~\ref{sec:pain points}, \ref{sec:expectations}, and \ref{sec:cost of migration}. Drawing from this evidence, we present two major conclusions that refine and expand our original hypothesis (Section \ref{sec:motivation}).

\vspace{1mm}
\noindent\textbf{Complexity Despite Technology Advancements}
Although cloud technologies have continued to evolve, the interview clearly shows that both cloud users, especially those in non-technical domains, still struggle with the complexity of real-world adoption. Instead of simplifying the development and deployment process, many modern cloud paradigms introduce new layers of abstraction that require even more specialized knowledge and operational overhead. These findings validate the original hypothesis that current cloud-native practices remain complicated, but they go further: the complexity is not simply due to immature technology but is often embedded in the very design of modern cloud platforms. This insight underscores the pressing need for a unified and streamlined approach to cloud infrastructure that can mitigate fragmentation and lower the technical barrier to adoption.

\vspace{1mm}
\noindent\textbf{Prioritizing Productivity and QoS over Cost and Migration}
In contrast to common assumptions that practitioners primarily consider cost and migration concerns when adopting new business-enabling technologies, our participants revealed a different reality. Across their roles and domains, neither cost nor migration emerged as the primary barrier to adoption. Instead, their focus centered on expectations such as Quality of Service (QoS), simplified programmability, and improved accessibility. In fact, many participants expressed a willingness to endure short-term migration challenges or higher upfront costs if the platform could offer meaningful improvements in service quality, operational simplicity, and development speed. This finding refines our original hypothesis by highlighting that while high cost is a consequence of the complicated solution life-cycle, the root cause is deeper, stemming from productivity gaps and architectural complexity in today’s cloud ecosystems.


\subsection{Domain-Specific Technology Adoption Patterns}

Analysis of our interview data reveals distinct technology adoption patterns across business domains, reflecting how sector-specific requirements and constraints shape infrastructure choices.

\vspace{1mm}
\noindent\textbf{Regulated Sectors Favor Hybrid and Private Cloud Architectures}

Organizations in heavily regulated domains---particularly finance and healthcare---demonstrate a strong preference for hybrid or private cloud deployments. Among finance participants (3 total), all three operate hybrid architectures where sensitive data remains on-premises while computation occurs in public cloud environments. This pattern is vividly illustrated in Case 4 (Finance Sector), where the participant explicitly noted the strict separation between on-premise data storage and encrypted cloud processing. Similarly, government-affiliated educational institutions (Case 5) cited privacy regulations (e.g., PDPA compliance) as prohibitive barriers to public cloud adoption.

This pattern reflects stringent data sovereignty, residency, and compliance requirements that cannot be easily satisfied by standard public cloud offerings. The technical consequence is increased \textit{development complexity} and \textit{operational maintainability} burden, as organizations must manage data movement, encryption pipelines, and compliance verification across hybrid boundaries.

\vspace{1mm}
\noindent\textbf{Edge Computing Concentrated in IoT and Industrial Applications}

Edge computing adoption is heavily concentrated in domains where physical proximity to data sources and network unreliability are critical concerns. Among IoT participants (3 total), all three deploy edge infrastructure. Similarly, all industrial sector participants (oil \& gas, manufacturing: 3 total) utilize edge deployments, often in hybrid edge-cloud configurations.

Representative use cases include: livestock monitoring systems in agriculture requiring local data collection due to remote farm connectivity; oil rig monitoring generating 1 MB/second of telemetry data with unreliable cellular networks (Case 2); and wearable air quality sensors in IoT applications prioritizing real-time responsiveness and data privacy. These scenarios share common challenges: unreliable network connectivity, latency-sensitive processing requirements, and the need for graceful degradation when cloud connectivity is unavailable.

Notably, 75\% (6 of 8) of edge deployments are hybrid edge-cloud architectures rather than pure edge solutions, indicating that edge computing typically augments rather than replaces cloud infrastructure. This hybrid approach balances local responsiveness with cloud-based analytics, storage, and management capabilities.

\vspace{1mm}
\noindent\textbf{Serverless Adoption Skews Toward Technical Organizations}

FaaS and serverless technologies exhibit uneven adoption across our tech/non-tech categorization. Among tech-oriented organizations (8 total), 62.5\% (5 participants) utilize serverless computing. In contrast, only 38\% (5 of 13) of non-tech organizations have adopted FaaS.

This disparity likely stems from serverless complexity: while FaaS promises infrastructure abstraction, it introduces conceptual challenges around stateless function design, cold start latency, orchestration, and debugging distributed workflows. Organizations with strong technical capabilities can navigate these trade-offs, whereas non-tech organizations often lack the expertise to effectively leverage serverless paradigms.

Interestingly, the finance sector shows high serverless adoption despite regulatory constraints, with 2 of 3 finance participants using FaaS for specific workloads (e.g., real-time ad generation, transaction processing). This suggests that when properly architected, serverless can coexist with strict compliance requirements.

\vspace{1mm}
\noindent\textbf{Multi-Cloud Strategies Remain Niche}

Despite industry discussions around vendor lock-in, only 19\% (4 of 21) of participants employ multi-cloud strategies. These adopters are concentrated in large enterprises (finance, industrial sectors) with sufficient resources to manage the operational complexity of multi-cloud deployments.

The limited adoption suggests that multi-cloud's theoretical benefits---vendor negotiation leverage, redundancy, best-of-breed service selection---are outweighed by practical concerns: increased operational overhead, skill set fragmentation, and challenges in maintaining consistent security, monitoring, and deployment workflows across platforms.

\vspace{1mm}
\noindent\textbf{IaaS Persistence Across All Domains}

Traditional IaaS remains prevalent across all domains (62\% overall adoption), including among tech organizations that might be expected to favor higher-level abstractions. This persistence reflects several factors: legacy application compatibility, desire for low-level infrastructure control, and maturity of IaaS tooling and operational practices.

Notably, several participants described hybrid IaaS/PaaS or IaaS/FaaS architectures where different service tiers address different workload characteristics. For example, database-intensive applications remain on IaaS for performance predictability, while stateless microservices leverage serverless for auto-scaling benefits. This pattern suggests that platform choice is increasingly workload-specific rather than organization-wide.

\subsection{Related Work}

Several studies have sought to clarify the challenges associated with cloud-native software development across various business contexts. For example, Oyeniran et al. \cite{oyeniran2024comprehensive} conducted a comprehensive review of how cloud-native technologies can enhance scalability and resilience in software development—two concerns frequently cited in both industry and academic discourse. Alka et al. \cite{alka2025entrepreneurial} examined the characteristics, impact, and challenges of cloud-native technologies in the context of entrepreneurship. Their study emphasized how the emergence of cloud-native paradigms has transformed the entrepreneurial landscape and introduced new opportunities, while also identifying nine major challenges specific to this domain. Deng et al. \cite{deng2024cloud} offered a systematic analysis of four core research domains in cloud-native application development—building, orchestration, operation, and maintenance—highlighting key service-oriented challenges that persist across the cloud-native life cycle.

Although these studies offer valuable insights into the challenges of cloud-native software development across business domains, they primarily rely on literature reviews and academic assumptions, with limited empirical validation grounded in practitioner experience. As a result, they overlook real-world constraints such as organizational readiness, skill shortages, integration complexity, and domain-specific expectations—factors that are often decisive in practical adoption scenarios. In contrast, our work is based on direct interviews with professionals across both technical and non-technical domains, providing a more grounded and context-aware understanding of the pain points, expectations, and migration concerns that organizations face.

\subsection{Potential Solutions}

To meet the demands of future cloud platforms that must deliver both quality of service and simplicity to support a broad spectrum of users, recent research has increasingly focused on SLA-aware abstractions and user-friendly deployment models.

Contemporary edge-cloud platforms emphasize QoS by supporting the specification of SLAs, typically expressed in terms of latency, availability, and cost. These SLAs guide the automated deployment and resource management of FaaS functions across heterogeneous infrastructures~\cite{goudarzi2022scheduling, nguyen2023storm}. For example, SLA-ORECS~\cite{lan2024sla} dynamically allocates resources to meet diverse SLA constraints, abstracting away the complexity of infrastructure decisions. A large body of research further enhances this direction by formulating SLA-driven placement and scheduling as optimization problems to balance user-defined QoS with cost efficiency~\cite{russo2024qos, baresi2024neptune, jarachanthan2022astrea}.

Many serverless platforms integrate functions and state data into cohesive deployment units, such as ``actors''~\cite{spenger2024survey} or ``processes''~\cite{copik2024process}, thereby simplifying the deployment of functions alongside their related state. At the forefront of this trend, OaaS and related works~\cite{lertpongrujikorn2023object,lertpongrujikorn2024object,lertpongrujikorn2024tutorial,lertpongrujikorn2024streamlining}, leverage object-oriented principles to encapsulate functions, state, and non-functional requirements within unified cloud deployment units.


\section{Summary}
\label{sec:summary}

This study, based on 21 in-depth interviews with professionals in both tech and non-tech sectors, confirms that complexity remains a significant barrier in cloud-native application development. This complexity is often embedded within the architecture of modern platforms. Rather than primarily being driven by cost or migration issues, participants consistently highlighted the need for improved productivity, quality of service (QoS), and developer accessibility. These insights challenge traditional assumptions and reveal a disconnect between current cloud-native practices and the practical needs of users. 

Ultimately, the findings underscore the urgent need to rethink how cloud infrastructure is designed, focusing on reducing abstraction overhead and accommodating diverse user capabilities. Future research should explore operational-simplified, SLA-aware, and user-friendly abstractions that strike a balance between service guarantees and simplicity. This includes designing end-to-end models that simplify the entire application life cycle—from development and deployment to monitoring and scaling.

\bibliographystyle{IEEEtran}
\balance
\bibliography{references}

\end{document}